\documentclass[aps,showpacs,twocolumn,amsmath,amssymb,floatfix,
nofootinbib,preprintnumbers]{revtex4}
\usepackage{epsfig}
\usepackage{graphicx}
\usepackage{amsmath}
\usepackage{color}
\usepackage{hyperref}
\newcommand{\be}{\begin{equation}}
\newcommand{\ee}{\end{equation}}
\newcommand{\beq}{\begin{equation}}
\newcommand{\eeq}{\end{equation}}
\newcommand{\bea}{\begin{eqnarray}}
\newcommand{\eea}{\end{eqnarray}}

\newcommand{\gev}{\, \text{GeV}}

\newcommand{\tin}{t_{\text{in}}}

\renewcommand{\Im}{\text{Im}\,}

\newcommand{\bsp}{\begin{sloppypar}}
\newcommand{\esp}{\end{sloppypar}}

\begin{document}


\title{Model-independent constraints on hadronic form factors with
above-threshold poles}
\author{Irinel Caprini}
\affiliation{Horia Hulubei National Institute for Physics and Nuclear
Engineering, POB MG-6, 077125 Bucharest-Magurele, Romania}
\author{Benjam\'{i}n Grinstein}
\affiliation{Department of Physics, University of California, San
Diego, La Jolla, California 92093, USA}
\author{Richard F. Lebed}
\affiliation{Department of Physics, Arizona State University, Tempe,
Arizona 85287-1504, USA}

\begin{abstract}\vspace{0.5cm} {Model-independent constraints on
    hadronic form factors, in particular those describing exclusive
    semileptonic decays, can be derived from the knowledge of field
    correlators calculated in perturbative QCD, using analyticity and
    unitarity.  The location of poles corresponding to below-threshold
    resonances, {\it i.e.}, stable states that cannot decay into a
    pair of hadrons from the crossed channel of the form factor, must
    be known {\it a priori}, and their effect, accounted for through
    the use of Blaschke factors, is to reduce the strength of the
    constraints in the semileptonic region.  By contrast,
    above-threshold resonances appear as poles on unphysical Riemann
    sheets, and their presence does not affect the original
    model-independent constraints.  We discuss the possibility that
    the above-threshold poles can provide indirect information on the
    form factors on the first Riemann sheet, either through
    information from their residues or by constraining the
    discontinuity function.  The bounds on form factors can be
    improved by imposing, in an exact way, the additional information
    in the extremal problem.  The semileptonic $K\to \pi\ell \nu$ and
    $D\to \pi\ell\nu$ decays are considered as illustrations.}
\end{abstract}

\maketitle

\section{Introduction}\label{sec:intro}
  
Since the pioneering works of Meiman \cite{Meiman:1963} and Okubo
\cite{Okubo:1971jf,Okubo:1971my}, it has been known that nontrivial
constraints on hadronic form factors can be derived from the knowledge
of suitably related field correlators.  The method was reconsidered
in \cite{Bourrely:1980gp} within the modern theory of strong
interactions, where the correlators relevant for the bounds on the
$K_{\ell 3}$ form factors were evaluated in the deep Euclidean region
by using perturbative QCD.

The method exploits unitarity and positivity of the spectral function,
and converts a dispersion relation for a correlator of two currents
into an integral condition along the unitarity cut ({\it i.e.}, above
the lowest production threshold of particles coupled to the currents)
for the modulus-square of the form factors parametrizing the relevant
matrix elements in the unitarity sum.  From this condition and the
analyticity properties of the form factors as functions of energy, one
can derive, with standard techniques of complex analysis \cite{Duren,
  KrNu}, constraints on the values of the form factors and their
derivatives at points inside the analyticity domain.

Many applications of this approach to heavy-quark form factors
describing $B\to D^{(*)}\ell\nu $ semileptonic decays, to
heavy-to-light form factors involved in $B\to \pi\ell\nu$ or $D\to
\pi\ell\nu$ decays, or to the light-meson form factors, have been
performed in the last 20 years
\cite{deRafael:1992tu, Carlson:1992kn, Falk:1992gw, Korner:1992hm,
Grinstein:1992hq, Caprini:1994fh, Caprini:1994np, Boyd:1994tt,
Boyd:1995sq, Boyd:1997qw, Boyd:1997kz, Lellouch:1995yv,
Caprini:1997mu, Caprini:1999ws, Bourrely:2005hp, Hill:2006bq,
Becher:2005bg, Bourrely:2008za, Ananthanarayan:2011uc, Abbas:2010ns}
(for a review of earlier literature see
\cite{Abbas:2010jc}).  A similar formalism has been applied also to
the electromagnetic form factors of the pion
\cite{Ananthanarayan:2011xt, Ananthanarayan:2012tn,
Ananthanarayan:2013zua, Ananthanarayan:2016mns} and proton
\cite{Hill:2010yb}, to the $\pi\omega$ form factor
\cite{Ananthanarayan:2014pta, Caprini:2015wja}, and to heavy baryons
\cite{Boyd:1995tg}.

The presence of singularities below the unitarity threshold modifies
the derivation of the bounds.  The method can be adapted to include in
an exact way the discontinuity across an unphysical cut below the
unitarity branch point, present in some cases, related to lighter
particles that can couple to the current \cite{Ananthanarayan:2014pta,
Caprini:2015wja}.  A pole situated below the unitarity threshold, of
known position but unknown residue, can be also accounted for in an
optimal way with the technique of Blaschke factors
\cite{Caprini:1994fh, Caprini:1994np}.  In such a case, the presence
of the pole leads to a weakening of the constraints.
 
Recently, the possible effect of resonances situated above the
unitarity threshold, close to the physical region, was discussed
in \cite{Grinstein:2015wqa}.  As known from general principles of
quantum field theory \cite{Peierls}, the unstable particles are
associated to complex poles in the energy plane.  Such poles cannot
appear on the first Riemann sheet of the complex plane, and instead
are situated on the second or higher Riemann sheets.  The argument
used in \cite{Grinstein:2015wqa} was based on the remark that a
complex pole on the second sheet close to the real axis produces a
local increase of the modulus of the form factor on the unitarity cut.
The same increase can be obtained however with a complex singularity
of the same position, but situated on the first Riemann sheet.
Therefore, in \cite{Grinstein:2015wqa} it was argued that the effect
of an above-threshold singularity can be mimicked through a complex
pole on the first Riemann sheet, near the physical region.  The latter
can be treated with the standard technique of Blaschke factors, much
like the subthreshold poles.  In this way,
Ref.~\cite{Grinstein:2015wqa} estimated the physical effect of the
presence of above-threshold resonances.

In the present paper, we consider the question whether the form-factor
parametrizations can be improved if some knowledge on the
above-threshold poles is provided.  We start with a brief review of
the technique of model-independent constraints, presenting in
particular the stronger constraints obtained with some additional
information outside the semileptonic decay range.  In
Sec.~\ref{sec:disc} we argue first that the presence of an
above-threshold pole does not affect the original bounds in the
semileptonic region. Then we investigate whether, from the presence of
an above-threshold resonance, one can obtain some information on the
form factor on the physical sheet and show that, in some cases, the
bounds can be improved by implementing additional information of this
type.  In particular, we find that the most practical constraints
arise from mapping the effect of the above-threshold resonance to the
phase of the form factor along the cut.  Our conclusions are given in
the last section.  In a short Appendix, we discuss the connection
between the first two Riemann sheets and the canonical variable $z$
used for solving the extremal problem.

\section{Model-independent constraints on hadronic form factors}
\label{sec:review}

We present below, following the review \cite{Abbas:2010jc} and the
recent paper \cite{Grinstein:2015wqa}, the main steps relevant for the
derivation of constraints on the form factor parametrizations.  As in
\cite{Grinstein:2015wqa}, we concentrate in particular on the form
factors relevant for the semileptonic decays of pseudoscalar mesons.
We consider the heavy-to-light ($Q \to q$) vectorlike ($V$, $A$, or $V
\! - \! A$) quark-transition current
\begin{equation} \label{eq:J}
J^\mu \equiv \bar Q \Gamma^\mu q \, ,
\end{equation}
and the two-point momentum-space Green's function $\Pi_J^{\mu \nu}$
separated into manifestly spin-1 ($\Pi_J^T$) and spin-0 ($\Pi_J^L$)
terms:
\begin{eqnarray}
&&\Pi_J^{\mu\nu} (q)  \equiv  i \! \int \! d^4 \! x \, e^{iqx} \left<0
\left| T J^\mu (x) J^{\dagger \nu} (0) \right| 0 \right> \nonumber \\ 
 &&=  \frac{1}{q^2} \left( q^\mu q^\nu - q^2
  g^{\mu\nu} \right) \Pi^T_J (q^2) + \frac{q^\mu q^\nu}{q^2} \Pi^L_J
(q^2) \, .
\label{eq:corr}
\end{eqnarray}
The functions $\Pi^{T,L}_J$ satisfy dispersion relations with positive
spectral functions, expressed by unitarity in terms of contributions
from a complete set of hadronic states.  From the asymptotic behavior
predicted by perturbative QCD, it follows that the dispersion
relations require subtractions (one for $\Pi^L_J$ and two for
$\Pi^T_J$).  The subtraction constants disappear by taking the
derivatives:
\begin{eqnarray}
&&\chi^L_J (q^2) \equiv \frac{\partial \Pi^L_J}{\partial q^2}  = 
\frac{1}{\pi} \int_0^\infty \! dt \, \frac{{\rm Im} \,
  \Pi^L_J(t)}{(t-q^2)^2} \, , \nonumber \\
&&\chi^T_J (q^2) \equiv \frac 1 2 \frac{\partial^2 \Pi^T_J}{\partial
  (q^2)^2}  =  \frac{1}{\pi} \int_0^\infty \! dt \, \frac{{\rm Im}
  \, \Pi^T_J(t)}{(t-q^2)^3} \, . \label{eq:DR}
\end{eqnarray}
Perturbative QCD can be used to compute the functions
$\chi^{\vphantom\dagger}_J(q^2)$ at values of $q^2$ far from the
region where the current $J$ can produce manifestly nonperturbative
effects like pairs of hadrons.  For heavy quarks, $Q = c$ or $b$, a
reasonable choice is $q^2 = 0$, while for $Q = s$ a spacelike value,
like $q^2 = -1 \gev^2$ or $q^2 = -2 \gev^2$, is necessary.

The spectral functions ${\rm Im} \, \Pi_J$ are evaluated by unitarity,
inserting into the unitarity sum a complete set of states $X$ that
couple the current $J$ to the vacuum:
\begin{equation} \label{eq:unitarity}
{\rm Im} \, \Pi^{\mu\nu}_J (q^2) = \frac{1}{2} \sum_X (2\pi)^4
\delta^4 (q - p_X)  \left< 0 \left| J^\mu \right| \! X \right> 
\left<  X \! \left| J^{\dagger\nu} \right|0 \right>  .
\end{equation}
For our purpose, it is enough to take $X$ to be the lightest meson
pair in which one of them (of mass $M$) contains a $Q$ quark and the
other (of mass $m$) contains a $\bar q$, and use the positivity of the
higher-mass contributions.  This choice gives a rigorous lower bound
on the spectral functions, in terms of the vector or scalar form
factors that parametrize the matrix elements of the current.  Using
the standard notation
\beq \label{eq:tpm}
t_\pm \equiv (M \pm m)^2 \, ,
\eeq
the inequality for the transverse polarization $\Pi^T_J$ can be
written as
\begin{equation} \label{eq:Fbd1}
\frac{1}{\pi \chi^T_J (q^2)} \int_{t_+}^\infty \! dt \frac{w(t) \,
|F(t)|^2}{(t-q^2)^3} \leq 1 \, ,
\end{equation}
where $t_+$ is the unitarity threshold, $F(t)$ is the vector form
factor, and $w(t)$ is a simple, nonnegative function, expressed as a
product of phase-space factors depending upon $t_+$ and $t_-$.  An
analogous expression holds for $\Pi^L_J$ and the scalar form factor.

Using the standard dispersion techniques in quantum field theory
\cite{Barton}, one can prove that the semileptonic form factors are in
general analytic functions in the complex $t$ plane, with a unitarity
cut along the real axis from $t_+$ to $\infty$.  In some cases, as in
$B\to D\ell\nu $ and $B\to \pi\ell\nu$, the form factors may also
exhibit poles situated on the real axis below the unitarity threshold
$t_+$.  No analogous poles are present in the form factors relevant in
$K\to \pi\ell\nu $ and $D\to \pi\ell\nu$ decays.  All the form factors
in semileptonic decays satisfy in addition the Schwarz reflection
condition, written generically as $F(t^*)=F^*(t)$.  The form factors
are therefore real on the real $t$ axis below $t_+$, in particular in
the semileptonic region $0\leq t\leq t_-$, where they can be measured
from the decay rates.

As shown in the pioneering papers \cite{Meiman:1963, Okubo:1971jf,
Okubo:1971my, Bourrely:1980gp}, one can obtain constraints on the
form-factor parametrizations in the semileptonic region, using their
analyticity properties and the boundary condition (\ref{eq:Fbd1}).  In
order to exploit this condition, it is convenient to map the cut $t$
plane onto the unit disk in the complex $z$ plane defined by the
conformal mapping\footnote{This definition differs by a minus sign
from that adopted in \cite{Grinstein:2015wqa}.}
\beq \label{eq:zdef} z\equiv \tilde z(t;t_0) \equiv \frac{\sqrt{t_+ -
    t_0} - \sqrt{t_+ - t}} {\sqrt{t_+ - t_0} + \sqrt{t_+ - t}}\, ,
\eeq
which maps the cut $t$ complex plane onto the interior of the unit
disk, such that the branch point $t_+$ in mapped onto $z=1$ and the
two edges of the unitarity cut $t \ge t_+$ map to the boundary
$|z|=1$.  Moreover, $z$ is real for $t \le t_+$.  The choice of the
free parameter $t_0$ in (\ref{eq:zdef}), which represents the point
mapped onto the origin of the $z$ plane, $\tilde z(t_0;t_0)=0$, will
be discussed below.

In the variable $z$, the inequality (\ref{eq:Fbd1}) is written in the
equivalent form
\beq \label{eq:FFz}
\frac{1}{2\pi i} \oint_C \frac{dz}{z} | \phi(z) F[\tilde t(z; t_0)]
|^2 \le 1 \,, \eeq
where 
\beq\label{eq:inv}
\tilde t(z; t_0)=\frac{4 z t_+ +t_0(1-z)^2}{(1+z)^2}
\eeq
is the inverse of (\ref{eq:zdef}), and $\phi(z)$ is an {\it outer
function\/}, defined in complex analysis \cite{Duren} as an analytic
function lacking zeros in $|z|<1$.  In our case, the function
$\phi(z)$ is defined by specifying its modulus
\beq \label{eq:outer}
|\phi(z)|^2 = \ \frac{w[\tilde t(z; t_0)]}{|d\tilde z(t;t_0)/dt| \,
  \chi^T (q^2) [\tilde t(z; t_0)-q^2]^3} \, ,
\eeq
on the boundary $z=e^{i\theta}$ of the unit disk.  Then the function
for $|z|<1$ can be reconstructed from its modulus on the boundary by
the representation \cite{Duren}
\beq\label{eq:phi}
\phi(z)=\exp\left[\frac{1}{2\pi} \int_{0}^{2\pi} {\rm d}\theta \,
\frac{e^{i\theta}+z}{e^{i\theta}-z}\,\ln |\phi(e^{i\theta})| \right].
\eeq
In particular cases of physical interest, $\phi(z)$ can be obtained in
closed form, as a product of simple analytic functions (see
\cite{Abbas:2010jc, Grinstein:2015wqa}).

From the boundary condition (\ref{eq:FFz}), one can derive constraints
on the form factor $F(t)$ at points inside the analyticity domain, in
particular in the semileptonic region.  It is important to emphasize
that the use of the outer function in (\ref{eq:FFz}) ensures the
constraints are optimal.  Assume first that the form factor $F(t)$ has
no singularities below the unitarity threshold $t_+$, being an
analytic function of real type [$F^*(t)=F(t^*)$] in the cut $t$ plane,
or equivalently in the unit disk $|z|<1$ (as mentioned above, this is
the case for the $K\pi$ or $D\pi$ form factors).  Then, expanding as:
\begin{equation} \label{eq:param}
  F(z) \equiv F[\tilde t(z;t_0)]=\frac{1}{ \phi(z)} \sum_{k=0}^\infty
  a_k z^k \, ,
\end{equation}
where the coefficients $a_k$ are real, the condition (\ref{eq:FFz})
reads:
\begin{equation} \label{eq:coef}
\sum_{k=0}^\infty a_k^2 \leq 1 \, .
\end{equation}
This inequality, which is valid also for any finite sum of terms, was
used in many studies to strongly constrain the parameters used in the
fits to semileptonic data or for estimating the truncation error
\cite{Boyd:1995sq, Boyd:1997qw, Boyd:1997kz, Caprini:1997mu,
Bourrely:2005hp, Hill:2006bq, Becher:2005bg, Bourrely:2008za}.  As
discussed in several papers, the truncation error is minimized by
choosing the parameter $t_0$ such that the semileptonic range $0\leq
t\leq t_- $ is mapped onto an interval $(-z_{\rm max}, z_{\rm max})$
symmetric around the origin in the $z$ plane.  This method allowed a
high-precision determination of the elements $V_{us}, V_{cb}$, and
$V_{ub}$ of the CKM matrix from exclusive semileptonic decays.

The constraints on the Taylor series coefficients $a_k$ become
stronger if some additional information on the form factor outside the
semileptonic range is available.  The general condition involving an
arbitrary number of coefficients $a_k$ and the values of $F(z)$ at an
arbitrary number of points inside the unit disk\footnote{In complex
analysis, if instead of the $L^2$ norm (\ref{eq:FFz}) the boundary
condition is expressed by means of the $L^\infty$ norm, the problem is
known as a combined Schur-Carath\'eodory and Pick-Nevanlinna
interpolation problem \cite{Duren, KrNu}.} has been derived using
several methods and can be found in \cite{Abbas:2010jc}.

For the discussion in the next section, it is of interest to give the
form of the constraint when one knows the values of the form factor
$F(z)$ at two complex-conjugate points, which we denote as $z_p$ and
$z_p^*$, with $|z_p|<1$.  Since the functions satisfy the Schwarz
reflection property, one has $F(z_p^*)=F^*(z_p)$.  Using, as in
\cite{Abbas:2010jc}, the technique of Lagrange multipliers for imposing
the additional constraints at $z_p$ and $z_p^*$, a straightforward
calculation gives the inequality
\beq \label{eq:domain1}
\sum_{k = 0}^{K-1} a_k^2\leq 1- {\cal F}(z_p, \xi),
\eeq
where ${\cal F}$ is defined as
\bea \label{eq:Fcal}
{\cal F}(z_p, \xi)&\!\!=\!\! & \frac{2
  (1-|z_p|^2)^2
  |1-z_p^2|^2}{|z_p|^{4 K}(z_p-z_p^*)^2}\\
& \!\!\times\!\! &\left[{\text{Re}}\left(\frac{\xi^2 z_p^{*2
        K}}{1-z_p^{*2}}\right)-\frac{|\xi|^2
    |z_p|^{2K}}{1-|z_p|^2}\right]\,, \nonumber
\eea
in terms of the point $z_p$ and the complex quantity 
\beq\label{eq:xi}
\xi= \phi(z_p) F(z_p)- \sum_{k=0}^{K-1}a_k z_p^k.
\eeq
The inequality (\ref{eq:domain1}) defines an allowed domain for first
$K$ coefficients $a_k$ in terms of the input complex value $F(z_p)$
entering the variable $\xi$.  One can check from (\ref{eq:Fcal}) that
the function ${\cal F}$ is positive for $|z_p|<1$ and arbitrary values
of $a_k$ and $F(z_p)$.  Therefore, the domain defined by
(\ref{eq:domain1}) is smaller than that given by the condition
\beq\label{eq:domain0}
\sum_{k = 0}^{K-1} a_k^2\leq 1
\eeq
derived from (\ref{eq:coef}).  As expected, knowledge of the value
$F(z_p)$ improves the constraints on the parameters in the
semileptonic region.  We note, however, that the improvement is small
if the point $z_p$ is close to the boundary of the unit disk, since
${\cal F}$ is small for $|z_p|$ close to 1.

Another additional piece of information that can improve the
constraints is knowledge of the phase of the form factor along a part
of the unitarity cut.  In some cases, as for the pion electromagnetic
form factor or the $K_{\ell 3}$ form factors, the phase is related by
Fermi-Watson theorem \cite{Fermi:2008zz, Watson:1954uc} to the phase
shift of the corresponding elastic scattering amplitude, which is
known with precision, for instance from the solution of Roy
equations \cite{Roy:1971tc}.  In the present context (as discussed in
the next section), it is of interest to note that one can
approximately obtain the phase on a part on the cut using the mass and
width of a nearby resonance.

Using this information as an additional constraint leads to a modified
optimization problem, solved for the first time for the $K_{\ell 3}$
form factors in \cite{Micu:1973vp}.  Several generalizations have been
discussed more recently in \cite{Caprini:1999ws, Bourrely:2005hp,
Abbas:2010jc}.  For completeness, we give below the constraint on the
first $K$ coefficients $a_k$ when the phase $\arg F(t)$ is known on
the region $0\leq t\leq \tin$ (for the derivation, see Sec.~4 of the
review \cite{Abbas:2010jc}).

We denote by 
\beq\label{eq:zetain}
\zeta_{\rm in} \equiv \tilde z(\tin; t_0)= e^{ i\theta_{\rm in}}
\eeq
the image on the unit circle in the $z$ plane of the point
$\tin+i\epsilon$ situated on the upper edge of the cut [the point
$\tin-i\epsilon$ being mapped onto $\exp(-i\theta_{\rm in}$)].  Then
the domain allowed for the coefficients $a_k$ is given by
\beq\label{eq:domain2}
    \sum_{k = 0}^{K-1} a_k^2+  \frac{1}{\pi} \sum_{k = 0}^{K-1} a_k
\int\limits_{-\theta_{\rm in}}^{\theta_{\rm in}} 
{\rm d} \theta \, \lambda(\theta) 
\sin\left[k \theta - \Phi(\theta) \right]  \leq 1,
\eeq
where
\beq
 \Phi(\theta)=\arg[F(e^{i\theta})] +\arg[\phi(e^{i\theta})] \, ,
\eeq
and $\lambda(\theta)$ is the solution of the integral equation
\beq\label{eq:eq1}
 \sum_{k = 0}^{K-1} a_k \sin[k \theta - \Phi(\theta)] =\lambda(\theta)
 - \frac{1}{2\pi} \! \int\limits_{-\theta_{\rm in}}^{\theta_{\rm in}}
 {\rm d} \theta' \lambda(\theta') \, {\cal K}_{\Phi}(\theta, \theta'),
\eeq
for $\theta \in (-\theta_{\rm in}, \theta_{\rm in}) $, where the
kernel is defined as
\beq\label{eq:calK}
{\cal K}_{\Phi}(\theta, \theta') \equiv \frac{\sin[(K-1/2) (\theta -
\theta') - \Phi(\theta) +
\Phi(\theta^\prime)]}{\sin[(\theta-\theta^\prime)/2]} \, .
\eeq
The inequality (\ref{eq:domain2}) describes an allowed domain for
$a_k$ that is smaller than the original domain (\ref{eq:domain0}),
which represents the improvement introduced by knowledge of the phase
on a part of the unitarity cut.

In the above derivations, the crucial role was played by the fact that
the form factor is analytic in the cut $t$ plane.  As discussed above,
the form factors relevant for $K\to\pi\ell\nu$ and $D\to\pi\ell\nu$
decays do not have subthreshold singularities, while the form factors
involved in $B\to D^{(*)}\ell\nu$ and $B\to \pi\ell\nu$ decays have
subthreshold poles, corresponding to particles stable with respect to
strong decays into $\bar B D$ and $\bar B \pi$, respectively.

As remarked for the first time in \cite {Caprini:1994fh,
  Caprini:1994np}, it is possible to derive constraints on the form
factor even if the residue of the pole is not known.  Denoting by
$z_p$ the position of the pole in the $z$-variable, the inclusion of
the pole can be done in an optimal way with respect to the condition
(\ref{eq:FFz}) by using a so-called Blaschke factor \cite{Duren}:
\beq\label{eq:Blaschke}
B(z;z_p)\equiv \frac{z-z_p}{1- z z_p^*}\,,
\eeq
which is a function analytic in $|z|\leq 1$ that vanishes at $z=z_p$
and has modulus unity for $z$ on the unit circle:
\beq\label{eq:Blb}
|B(\zeta; z_p)|=1, \quad \quad \zeta=e^{i \theta}.
\eeq
By using (\ref{eq:Blb}), one obtains from (\ref{eq:FFz}), with no loss
of information, the equivalent condition
\begin{equation} \label{eq:FFz1}
  \frac{1}{2\pi i} \oint_C \frac{dz}{z} | B(z;z_p) \phi(z)
  F[\tilde t(z; t_0)] |^2 \le 1\,.
\eeq
Taking into account that the product $B(z;z_p) F(z)$ is analytic in
$|z|<1$, we write the most general parametrization of the form factor
as
\begin{equation} \label{eq:paramB}
F(z)=\frac{1}{B(z;z_p)\phi(z)} \sum_{k=0}^\infty a_k z^k\,,
\end{equation}
where the coefficients $a_k$ still satisfy (\ref{eq:coef}).
 
Since by the maximum modulus principle $|B(z; z_p)|<1$ for $|z|<1$,
the constraints in the semileptonic region derived from
(\ref{eq:paramB}) are weaker than those valid when no subthreshold
poles are present.

\section{Above-Threshold Poles}\label{sec:disc}

The possible effect of an above-threshold resonance was investigated
in \cite{Grinstein:2015wqa}, starting with the remark that a pole in
the form factor at the same position as the resonance pole, but
situated on the first Riemann sheet, creates a Breit-Wigner lineshape
indistinguishable from that created by a physical second-sheet pole
equally near the unitarity cut.  Therefore, the effect of a
second-sheet pole was simulated by a pole situated on the first sheet.
In Appendix A we give for completeness the positions in the $z$ plane
of a second-sheet pole, $z_p^{\rm II}$, and its counterpart on the
first sheet, $z_p^{\rm I}$, for some particular form factors.  The
treatment of the fake pole at $z_p\equiv z_p^{\rm I}$ by the technique
of Blaschke factors, as shown in the previous section, led to the
conclusion that an above-threshold resonance has the effect of
weakening the unitarity bounds.  The effect was found to be small in
the case of the $K\pi$ and $D\pi$ vector form factors.  However, since
any information on the modulus of the form factor on the cut is
covered by the rigorous condition (\ref{eq:Fbd1}), which is the main
ingredient of the formalism, one can see that accounting for the fake
pole is not necessary.  Thus, the presence of an above-threshold pole
does not affect the bounds in the semileptonic region.

On the other hand, it is known that a pole of the scattering amplitude
as a function of c.m.\ energy squared on a higher Riemann sheet can
produce in some cases (such as elastic $2 \to 2$ scattering
\cite{Barton}) a reflection on the first sheet.  Thus, a pole due to a
resonance on the second sheet induces a zero of the $S$-matrix element
at the corresponding point on the first sheet.  This property is
useful in practice: In \cite{Caprini:2005zr}, the mass and width of
the $\sigma$ scalar resonance were found by performing the analytic
continuation of the Roy equations for $\pi\pi$ scattering into the
first sheet of the complex plane and looking for the zeros of the $S$
matrix.

One might ask whether a similar property exists for form factors.  In
order to answer this question, we consider in more detail the analytic
continuation to the second Riemann sheet.  According to the
general dispersive approach in field theory \cite{Barton}, it is
useful to consider, along with a given form factor $F(t)$, the
corresponding  amplitude (of definite angular momentum and isospin) of
the elastic scattering of two hadrons of masses $M$ and $m$.  We
review below some well-known facts about these quantities that are
useful for our purpose.

Denoting by $f(t)$ the relevant partial wave of the invariant elastic
amplitude, elastic unitarity is expressed as
\beq\label{eq:unitT}
\Im f(t)=\rho(t) f(t) f^*(t) \, ,\quad\quad t_+\leq t \leq
\tin \, ,
\eeq
where $\rho(t)= \sqrt{(1-t_+/t)  (1-t_-/t)}$ is the dimensionless
phase space.  This relation is valid in the elastic region, below the
opening of the first inelastic threshold $\tin$.  
Unless otherwise specified,  by real $t$ above the
threshold $t_+$, we mean the value $t +i \epsilon$, on the upper edge
of the cut.

Equation (\ref{eq:unitT}) has the well-known solution  \cite{Barton}
\beq\label{eq:fdelta}
f(t)=\frac{e^{i\delta(t)} \sin \delta(t)}{\rho(t)}\, ,
\quad\quad t_+\leq t \leq \tin \, ,
\eeq
in terms of the phase shift $\delta(t)$.

The relation (\ref{eq:unitT}) provides also the route for analytic
continuation to the second Riemann sheet.  Using the Schwarz
reflection property $f^*(t)= f(t^*)$, we write (\ref{eq:unitT}) as
\beq\label{eq:unitT1}
  f(t+i\epsilon )-f(t-i\epsilon)= 2 i \rho(t) f(t+i\epsilon)
  f(t-i\epsilon) \, .
\eeq
The amplitude $f^{\rm II}(t)$ on the second sheet is defined by gluing
the lower edge of the cut in the first sheet to the upper edge on the
cut in the second sheet, {\em i.e.}, by requiring $f^{\rm
II}(t+i\epsilon)= f(t-i\epsilon)$.  Understanding all quantities
without a superscript as defined on the first Riemann sheet, we
write Eq.~(\ref{eq:unitT1}) as
\beq\label{eq:TII}
f^{\rm II}(t)=\frac{f(t)}{1+2 i \rho(t) f(t)} \, .
\eeq
The $S$ matrix is defined on the first sheet as
\beq\label{eq:S}
S(t)= 1+2 i \rho(t) f(t) \, ,
\eeq
and on the second sheet as 
\beq\label{eq:SII}
S^{\rm II}(t)=1 -2 i \rho(t) f^{\rm II}(t) \, .
\eeq
Using the definition (\ref{eq:TII}) of $f^{\rm II}(t)$, one obtains:
\beq\label{eq:SII1}
S^{\rm II}(t)=\frac{1}{S(t)} \, .
\eeq
From this relation it follows that the poles of $f^{\rm II}(t)$ [and
of $S^{\rm II}(t)$] correspond to zeros of $S(t)$ on the first sheet,
the property mentioned at the beginning of this section.

Turning now to form factors, elastic unitarity implies the relation
\cite{Barton}
\beq\label{eq:unitF}
 \Im F(t)= \rho(t) F^*(t) f(t)\,,
\eeq 
valid for $t$ in the elastic region, $t_+\leq t \leq \tin$.

A first consequence of (\ref{eq:unitF}) is the well-known Fermi-Watson
theorem \cite{Fermi:2008zz,Watson:1954uc}: Since the right-hand side
is known to be real, the phase of the form factor must be equal to the
phase shift of the amplitude (\ref{eq:fdelta}):
\beq\label{eq:FW} 
\arg [F(t)]=\arg [f(t)]=\delta(t)\,, \quad t_+\leq t \leq \tin\,.
\eeq
Moreover, by defining, in analogy to $f^{\rm II}(t)$,
\beq\label{eq:FIIdef}
F^{\rm II}(t+i\epsilon )\equiv F(t-i\epsilon ) \, ,
\eeq
one obtains from (\ref{eq:unitF}):
\beq\label{eq:FII}
   F^{\rm II}(t)=\frac{F(t)}{1+2 i \rho(t) f(t)}=\frac{F(t)}{S(t)}.
\eeq
Assuming that $F(t)$ does not vanish at the zero of $S(t)$, $F^{\rm
II}(t)$ has a pole at that position.  So, the second-sheet poles of
the form factor and the  $S$-matrix element have
the same position, a known universality property of the poles in
$S$-matrix theory.  The relation (\ref{eq:FII}) shows also that the
analytic structure of the function $F^{\rm II}(t)$ is more complicated
that that of $F(t)$: besides the unitarity cut, it has the same branch
points as $S(t)$, in particular those lying on the left-hand cut
produced by crossed-channel exchanges \cite{Barton}.

We show now that it is possible to express the value of $F(t)$ on the
first sheet, at the value of $t$ corresponding to the second-sheet
pole position, in terms of the residues of the poles of the form
factor and the amplitude on the second sheet.  From (\ref{eq:SII1})
and (\ref{eq:FII}) one has:
\beq\label{eq:FIdef}
  F(t)= F^{\rm II}(t) S(t)=\frac{F^{\rm II}(t)}{ S^{\rm II}(t)} \, .
\eeq 
Denoting by $t_p$ one of the pole positions on the second sheet, in
the vicinity of the pole one can write
\beq\label{eq:r1}
   f^{\rm II}(t) =\frac{ r_f}{t-t_p}+g(t) \, ,
\eeq
and
\beq\label{eq:r2}
    F^{\rm II}(t) = \frac{r_F}{t-t_p}+h(t) \, ,
\eeq
where the functions $g$ and $h$ are regular at $t=t_p$.
Using these expressions and (\ref{eq:SII}) in (\ref{eq:FIdef}) and
taking the limit $t\to t_p$ gives
\beq\label{eq:Ftp}
  F(t_p)=\frac{i}{2 \rho(t_p)}\ \frac{r_F} {r_f} \, .
\eeq
From the Schwarz principle, $F(t_p^*)= F^*(t_p)$, the value of $F$ at
$t_p^*$ (still on the first sheet) is the complex conjugate of the
expression (\ref{eq:Ftp}).  As shown in the previous section, this
additional condition on the first sheet can be included exactly in the
Meiman-Okubo problem, leading to an improvement of the bounds in the
semileptonic region.  The relation (\ref{eq:domain1}) gives the
allowed domain of the coefficients $a_k$ in terms of this additional
information.  It can be viewed therefore as a new sum rule relating
the residues of the above-threshold poles on the second Riemann sheet
to the parameters describing the semileptonic decays.

\begin{figure}[htb]\vspace{1cm}
\begin{center} \includegraphics[width = 8.5cm]{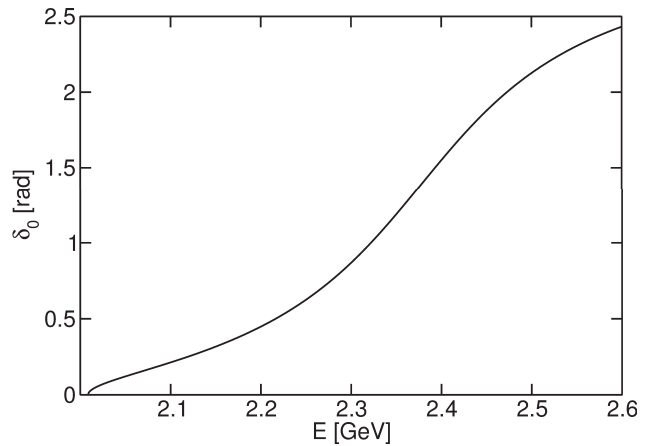}
\caption{Phase of the $D\pi$ scalar form factor as a
  function of the c.m.\ energy
  $E=\sqrt{t}$. \label{fig:phase}}\end{center}
\end{figure}
In practice, if the ratio $r_F/r_f$ is not known, one can reverse the
argument and use (\ref{eq:domain1}) as a constraint on the residues,
in terms of the coefficients $a_k$ determined from fits to
semileptonic decay data.  However, the correlation is expected to be
small, due to the fact that, as shown in Appendix A, in cases of
interest the point $z_p\equiv z_p^{\rm I}$ is close to the boundary
$|z|=1$.  Therefore, the value of the new sum rule in this case is of
more formal than phenomenological significance.

Of more practical value turns out to be another consequence of
unitarity that is valid on the unitarity cut below the first
inelastic threshold.  By dividing both sides of (\ref{eq:unitT1}) by
the product $f^* \! f$, one has
\beq
  \frac{1}{f^*(t)}-\frac{1}{f(t)}=2 i \rho(t) \, ,
\eeq
which implies
\beq\label{eq:Tinv}
  \Im \left[\frac{1}{f(t)}\right] = - \rho(t) \, .
\eeq
The solution of this equation is
\beq\label{eq:Tunit}
  f(t)=\frac{1}{\psi(t) - i \rho(t)} \, ,
\eeq
where the undetermined function $\psi(t)$ is real on the elastic part
of the unitarity cut, $t_+\leq t\leq \tin$.  If a narrow resonance of
mass $M$ and width $\Gamma$ is present, this function can be
parametrized  as
\beq\label{eq:psi}
\psi(t)\sim \frac{M^2-t}{M\Gamma}\,,
\eeq
up to factors holomorphic in a region $t_+<t<\tin$, where $\tin$
denotes the first inelastic threshold.  By including all these factors
in an energy-dependent $\Gamma(t)$, we can write, with a good
approximation, the phase of the form factor in a limited energy region
above the threshold as:
\beq\label{eq:phase}
\arg[F(t)]=\arctan\left[\frac{M \Gamma(t)}{M^2-t}\right]\,.
\eeq
This relation can be generalized to the case where overlapping
resonances occur.  In such a case, it is a well-known feature of
$S$-matrix theory that simply summing Breit-Wigner resonances does not
preserve unitarity, and the proper treatment would require allowing
$\Gamma$ not only to be dependent on energy, but also a matrix-valued
quantity over the various channels.

In Fig.~\ref{fig:phase} we show the phase $\delta_0$ of the scalar
$D\pi$ form factor obtained from (\ref{eq:phase}), using the standard
Breit-Wigner expression $\Gamma(t)=\Gamma \rho(t)/\rho(M^2)$, with the
mass $M= 2.351 \gev$ and width $\Gamma = 0.230 \gev $ of the $D_0^*$
resonance \cite{Olive:2016xmw}.  We can assume that this value of the
phase is a good approximation in the elastic region, below the opening
of inelastic channels.

As discussed in the previous section, this additional information
leads to a stronger constraint in the semileptonic region, given by
Eq.~(\ref{eq:domain2}).  This constraint can be easily derived by
solving the integral equation (\ref{eq:eq1}) for the function
$\lambda(\theta)$ and using this solution in (\ref{eq:domain2}).  For
illustration, we present below the result of this analysis for the
scalar $D\pi$ form factor.  We take the value $\chi^L_V(q^2=0)=0.016$
from Ref.~\cite{Ananthanarayan:2011uc} and the outer function from
Refs. \cite{Ananthanarayan:2011uc, Abbas:2010jc}:
\begin{eqnarray}
\phi(z) & = & \frac{\sqrt{3}}{32 \sqrt{\chi^L_V(0)}
\sqrt{\pi}}\frac{m_D-m_\pi} {m_D+m_\pi} (1-z)(1+z)^{3/2} \nonumber \\
& & \times \frac{ (1-z z_-)^{1/2}}{ (1+ z_-)^{1/2}} \, ,
\label{eq:outerDpi}
\end{eqnarray}
where we take for simplicity $t_0=0$ in (\ref{eq:zdef}) and use the
notation $z_-\equiv \tilde z(t_-;0)$.

Taking for illustration $K=5$, we obtain the allowed domain for the
coefficients $a_k$, $k\leq 4$:
\bea\label{eq:domain3}
&~&1.78 \, a_0^2 + 1.28 \, a_1^2+ 1.13 \, a_2^2+ 1.84\, a_3^2 + 2.33\,
a_4^2 \nonumber \\
&+& 0.79\, a_0 a_1  - 0.49\, a_0 a_2 - 1.61\, a_0 a_3- 1.96\,
a_0 a_4\nonumber \\
&-& 0.13\, a_1 a_2   - 0.84\, a_1 a_3   - 1.09\, a_1 a_4\nonumber\\
& +& 0.49\, a_2 a_3 +0.54\, a_2 a_4 + 2.09\, a_3 a_4 \ \leq \ 1.
\eea
In this calculation, we assume that the phase is given up to the first
inelastic threshold $\tin=(2.42\gev)^2$ due to the $D\eta$
channel. The results are actually quite stable against the variation
of $\tin$ around this value.

It is easy to see that the constraint (\ref{eq:domain3}) is stronger
than the standard condition (\ref{eq:domain0}).  In a typical
application to semileptonic processes, the lowest coefficients $a_k$
are determined from fits of the data, and the aim is to set a bound on
the next coefficient, which gives an estimate of the truncation
error. In practical applications (see for instance
\cite{Bourrely:2008za}), the optimal values of the parameters are
usually small, far from saturating the upper bound
(\ref{eq:domain0}). To simulate such a situation, we take, for
instance, the input values $a_0=0.10, \, a_1=0.08, \, a_2=0.07$ and
$a_3=0.05$, for which the left hand side of (\ref{eq:domain0}) is
0.024.  With this input, we obtain the constraint $|a_4|\leq 0.99$
from the standard inequality (\ref{eq:domain0}), and the smaller range
$-0.62\leq a_4\leq 0.68$ from the improved constraint
(\ref{eq:domain3}).  We can then obtain a bound on the truncation
error $\delta F(t_-)$ at the end $t_-$ (corresponding to $z_-$) of the
semileptonic region.  From the parametrization (\ref{eq:param}), one
can write this error as:
\beq\label{eq:error}
\delta F(t_-)\approx \frac{|a_4| z_-^4}{|\phi(z_-)|} \, .
\eeq
Using the above limits on $a_4$ and the values $z_-=0.325$ and
$\phi(z_-)=0.176$ in our case, we obtain from (\ref{eq:error}) the
uncertainties $\delta F(t_-)\approx 0.063$ using the standard
constraint (\ref{eq:domain0}) and $\delta F(t_-)\approx 0.043$ using
the improved constraint (\ref{eq:domain3}), which amounts to an
improvement by about 30\%. Similar results are obtained for a
large class of input values for the lowest coefficients.

One can use also the optimal value of $t_0$ discussed in
Sec.~\ref{sec:review}, for which the semileptonic region is mapped
onto a symmetric range in the $z$ plane. From Eq.~(\ref{eq:t0}), we
obtain in our case $t_0=1.97 \gev^2$ and $z_-=0.167$. Due to the
smaller $z_-$, the error estimated from (\ref{eq:error}) is much
smaller, but the constraints on the coefficient $a_4$ are similar to
those reported above. In this case too, the improvement brought by the
incorporation of the phase $\delta_0$ turns out to be quite important.

\section{Summary and conclusions}\label{sec:conc}

In this paper we have continued the discussion of the effect of
above-threshold singularities on model-independent form-factor
parametrizations, initiated in Ref.~\cite{Grinstein:2015wqa}.  We
emphasized the fact that the presence of above-threshold poles does
not affect the strength of the original model-independent
constraints. By exploiting the connection between the first and the
second Riemann sheets of a generic semileptonic form factor, we have
derived a relation between the value of the form factor on the first
Riemann sheet at the point $t_p$ that is the image of the location of
the resonance pole on the unphysical (second) Riemann sheet, and the
residues of the form factor and of the related elastic scattering
amplitude.  Using this expression in the combined constraint
(\ref{eq:domain1}) involving the coefficients $a_n$ of a Taylor series
expansion in the variable $z$ and the values of the form factor at the
two complex-conjugate points, we derived a new sum rule relating the
parametrization in the semileptonic region to the residues of the
second-sheet poles of the form factor $F$ and the corresponding
elastic scattering amplitude $f$. We argued however that the effect of
this additional information in improving the model-independent
constraints is expected to be small.  Finally, we showed that from the
mass and width of a narrow resonance, one can approximately obtain the
phase of the form factor on a limited part of the unitarity cut.  By
including this additional information in the extremal problem, one
obtains stronger constraints, given in (\ref{eq:domain2}), on the
form-factor parametrization in the semileptonic region. This second
method appears to be of more immediate utility in phenomenological
applications.

\vspace{0.4cm}

\subsection*{Acknowledgments} 
I.C.\ acknowledges support from the Ministry of Research and Innovation, Contract PN 16420101/2016.
B.G.\ was supported by the U.S.\ Department of Energy under Grant
DE-SC0009919.  R.F.L.\ was supported by the U.S.\ National Science
Foundation under Grant No.\ 1403891.

\appendix

\section{Uniformization of the two-sheet Riemann surface by $z$ mapping}

In this Appendix we discuss the connection between the canonical
variable $z$ in Eq.~(\ref{eq:zdef}) used for solving the extremal
problem in Sec.~\ref{sec:review} and the Riemann structure of the
elastic cut of the semileptonic form factor $F(t)$.  We first note
that (\ref{eq:zdef}) can be written as
\beq\label{eq:zdefk}
z\equiv \tilde z(t;t_0)=\frac{  \sqrt{t_+ - t_0} +  i k(t)}{
\sqrt{t_+ - t_0} -  i k(t)},
\eeq
in terms of the function
\beq\label{eq:k}
k(t)=\sqrt{t-t_+}\,.
\eeq
We recall that the first Riemann sheet is defined by $\arg (t-t_+)\in
(0, 2 \pi)$, while the second sheet is defined by $\arg (t-t_+)\in (2
\pi, 4\pi)$.  It follows that the first Riemann sheet corresponds to
$\arg k(t)\in (0, \pi)$, which implies $k_I(t) > 0$, and the second
Riemann sheet corresponds to $\arg k(t)\in (\pi, 2 \pi)$, which
implies $ k_I(t) < 0$, where $k_I(t)$ is the imaginary part of $k(t)$.
Denoting by $k_R(t)$ the real part of $k(t)$, we obtain from
(\ref{eq:zdefk}):
\beq\label{eq:zmod}
|z|^2=\frac{ [\sqrt{t_+ - t_0} - k_I(t)]^2 +  k^2_R(t)}
{ [\sqrt{t_+ - t_0} + k_I(t)]^2 +  k^2_R(t)}\,.
\eeq
From this relation it follows that 
\bea
&k_I(t)>0 \quad \Rightarrow \quad |z|<1, \nonumber\\
&k_I(t)<0 \quad \Rightarrow \quad |z|>1.
\eea
Therefore, the first Riemann sheet of the $t$ plane, where $k_I(t)>0$,
is mapped inside the unit circle in the $z$ plane, while the second
sheet, where $k_I(t)<0$, is mapped outside the unit circle.  In
standard terminology, the variable (\ref{eq:zdef}) achieves the
uniformization of the Riemann surface of the elastic cut, {\em i.e.},
it maps the two Riemann sheets onto a single plane.

For the discussion in Sec.~\ref{sec:disc}, it is useful to have a
relation between the images in the $z$ plane of the pole on the second
sheet, and of the corresponding complex point situated on the first
sheet.  This relation follows from the symmetry property
\beq\label{eq:symm}
\tilde t(z; t_0)=\tilde t(z^{-1}; t_0) \, ,
\eeq
satisfied by (\ref{eq:inv}), which shows that the images in the $z$
plane of the first-sheet and second-sheet points corresponding to the
same complex $t$ value are inverse to each other,
\beq
z_p^{\rm I}=\frac{1}{z_p^{\rm II}}\,. \eeq

For a  numerical illustration,  we take for definiteness 
\begin{equation}\label{eq:t0}
t_0 = t_+ \left[ 1 - \sqrt{ 1 - \frac{t_-}{t_+} } \right] \, ,
\end{equation}
to achieve a symmetric semileptonic range ($-z_{\rm max}$, $z_{\rm
max}$), as discussed in Sec.~\ref{sec:review}.  Then, using the masses
and widths from \cite{Olive:2016xmw} for the poles associated with the
first vector resonances $K^*(892)$ and $D^*(2010)$ for the $K\pi$ and
$D\pi$ vector form factors, respectively, we obtain from
(\ref{eq:zdefk}) the positions in the $z$ plane of $z_p^{\rm II}$ and
their first-sheet counterparts $z_p^{\rm I}$.  They read:
\bea\label{eq:K*z}
&z_p^{\rm II}= -0.11 \mp 1.05 \,i, \quad \quad |z_p^{\rm II}|=1.06
\,,\\ &z_p^{\rm I}= -0.10 \pm 0.94 \,i, \quad \quad |z_p^{\rm I}|=
0.95\,,\nonumber
\eea
and   
\bea\label{eq:D*z}
&z_p^{\rm II}=0.978 \mp 0.212 \,i, \quad \quad |z_p^{\rm II}|=1.001
\,,\\ &z_p^{\rm I}= 0.977 \pm 0.212 \,i, \quad \quad |z_p^{\rm
I}|=0.999 \,,\nonumber
\eea
respectively.  For the scalar resonance $D^*_0(2400)$ relevant to the
scalar $D\pi$ form factor, the corresponding points are
\bea\label{eq:D0z}
&z_p^{\rm II}= 0.151 \mp 1.179\, i, \quad \quad |z_p^{\rm II}|=1.19
\,,\\ &z_p^{\rm I}= 0.107 \pm 0.834 \, i, \quad \quad |z_p^{\rm
I}|=0.84 \,.\nonumber
\eea
We emphasize that the form factors have poles at the points $z_p^{\rm
II}$, but are regular at $z_p^{\rm I}$.



\begin{thebibliography}{150}

\bibitem{Meiman:1963}
  N.N.~Meiman,
  ``Analytic expressions for upper limits of coupling constants in
  quantum field theory,''
  Zh.\ Eksp.\ Teor.\ Fiz.\ {\bf 44}, 1228 (1963) [Sov.\ Phys.\ JETP
  {\bf 17}, 830 (1963)].


\bibitem{Okubo:1971jf} 
  S.~Okubo,
  ``Exact bounds for $K_{\ell 3}$ decay parameters,''
  Phys.\ Rev.\ D {\bf 3}, 2807 (1971).

\bibitem{Okubo:1971my} 
  S.~Okubo,
  ``New improved bounds for $K_{\ell 3}$ parameters,''
  Phys.\ Rev.\ D {\bf 4}, 725 (1971).


\bibitem{Bourrely:1980gp} 
  C.~Bourrely, B.~Machet, and E.~de~Rafael,
  ``Semileptonic decays of pseudoscalar particles ($M \to M^\prime
  \ell \nu_\ell$) and short distance behavior of quantum
  chromodynamics,''
  Nucl.\ Phys.\ B {\bf 189}, 157 (1981).


\bibitem{Duren}
  P.L.~Duren, {\it Theory of $H^{\rm p}$ Spaces}, Academic Press, New
  York, 1970.

\bibitem{KrNu}
  M.G.~Krein and P.I.~Nudelman, ``On some new problems for functions
  of Hardy class and continual families of functions with double
  orthogonality,''
  Sov.\ Math.\ Dokl.\ {\bf 14}, 435 (1973)
  [Dokl.\ Acad.\ Nauk.\ SSSR {\bf 209}, 537 (1973)].

\bibitem{deRafael:1992tu} 
  E.~de~Rafael and J.~Taron,
  ``Constraints on heavy meson form-factors,''
  Phys.\ Lett.\ B {\bf 282}, 215 (1992).


\bibitem{Carlson:1992kn} 
  C.E.~Carlson, J.~Milana, N.~Isgur, T.~Mannel, and W.~Roberts,
  ``Comment regarding bounds upon heavy meson form-factors,''
  Phys.\ Lett.\ B {\bf 299}, 133 (1993).

\bibitem{Falk:1992gw} 
  A.F.~Falk, M.E.~Luke, and M.B.~Wise,
  ``Analyticity and the Isgur-Wise function,''
  Phys.\ Lett.\ B {\bf 299}, 123 (1993).

\bibitem{Korner:1992hm} 
  J.G.~K\"{o}rner, D.~Pirjol, and C.~Dominguez,
  ``Analyticity bounds on the Isgur-Wise function,''
  Phys.\ Lett.\ B {\bf 301}, 257 (1993).

\bibitem{Grinstein:1992hq} 
  B.~Grinstein and P.F.~Mende,
  ``On constraints for heavy meson form-factors,''
  Phys.\ Lett.\ B {\bf 299}, 127 (1993)
  [hep-ph/9211216].

\bibitem{Caprini:1994fh} 
  I.~Caprini,
  ``Effect of upsilon poles on the analyticity constraints for heavy
  meson form-factors,''
  Z.\ Phys.\ C {\bf 61}, 651 (1994).

\bibitem{Caprini:1994np} 
  I.~Caprini,
  ``Slope of the Isgur-Wise function from a QSSR constraint on the
  $\Upsilon B \bar B$ couplings,''
  Phys.\ Lett.\ B {\bf 339}, 187 (1994)
  [hep-ph/9408238].

\bibitem{Boyd:1994tt} 
  C.G.~Boyd, B.~Grinstein, and R.F.~Lebed,
  ``Constraints on form-factors for exclusive semileptonic heavy to
  light meson decays,''
  Phys.\ Rev.\ Lett.\  {\bf 74}, 4603 (1995)
  [hep-ph/9412324].

\bibitem{Boyd:1995sq} 
  C.G.~Boyd, B.~Grinstein, and R.F.~Lebed,
  ``Model independent determinations of $\bar B \to D \ell \bar \nu$,
  $D^* \ell \bar \nu$ form-factors,''
  Nucl.\ Phys.\ B {\bf 461}, 493 (1996)
  [hep-ph/9508211].


\bibitem{Boyd:1997qw} 
  C.G.~Boyd and M.J.~Savage,
  ``Analyticity, shapes of semileptonic form-factors, and $\bar B \to
  \pi \ell \bar \nu$,''
  Phys.\ Rev.\ D {\bf 56}, 303 (1997)
  [hep-ph/9702300].

\bibitem{Boyd:1997kz} 
  C.G.~Boyd, B.~Grinstein, and R.F.~Lebed,
  ``Precision corrections to dispersive bounds on form-factors,''
  Phys.\ Rev.\ D {\bf 56}, 6895 (1997)
  [hep-ph/9705252].

\bibitem{Lellouch:1995yv} 
  L.~Lellouch,
  ``Lattice constrained unitarity bounds for $\bar B^0 \to \pi^+
  \ell^- \bar \nu_\ell$ decays,''
  Nucl.\ Phys.\ B {\bf 479}, 353 (1996)
  [hep-ph/9509358].


\bibitem{Caprini:1997mu} 
  I.~Caprini, L.~Lellouch, and M.~Neubert,
  ``Dispersive bounds on the shape of $\bar B \to D^{(*)} \ell \bar
  \nu$ form-factors,''
  Nucl.\ Phys.\ B {\bf 530}, 153 (1998)
  [hep-ph/9712417].


\bibitem{Caprini:1999ws} 
  I.~Caprini,
  ``Dispersive and chiral symmetry constraints on the light meson
  form-factors,''
  Eur.\ Phys.\ J.\ C {\bf 13}, 471 (2000)
  [hep-ph/9907227].


\bibitem{Bourrely:2005hp} 
  C.~Bourrely and I.~Caprini,
  ``Bounds on the slope and the curvature of the scalar $K \pi$
  form-factor at zero momentum transfer,''
  Nucl.\ Phys.\ B {\bf 722}, 149 (2005)
  [hep-ph/0504016].


\bibitem{Hill:2006bq} 
  R.J.~Hill,
  ``Constraints on the form factors for $K \to \pi \ell \nu$ and
  implications for $|V_{us}|$,''
  Phys.\ Rev.\ D {\bf 74}, 096006 (2006)
  [hep-ph/0607108].


\bibitem{Becher:2005bg} 
  T.~Becher and R.J.~Hill,
  ``Comment on form-factor shape and extraction of $|V_{ub}|$ from $B
  \to \pi \ell \nu$,''
  Phys.\ Lett.\ B {\bf 633}, 61 (2006)
  [hep-ph/0509090].


\bibitem{Bourrely:2008za} 
  C.~Bourrely, I.~Caprini, and L.~Lellouch,
  ``Model-independent description of $B \to \pi \ell \nu$ decays and
  a determination of $|V_{ub}|$,''
  Phys.\ Rev.\ D {\bf 79}, 013008 (2009)
  [Erratum: Phys.\ Rev.\ D {\bf 82}, 099902 (2010)]
  [arXiv:0807.2722 [hep-ph]].


\bibitem{Ananthanarayan:2011uc} 
  B.~Ananthanarayan, I.~Caprini, and I.~Sentitemsu Imsong,
  ``Implications of unitarity and analyticity for the $D\pi$ form
  factors,''
  Eur.\ Phys.\ J.\ A {\bf 47}, 147 (2011)
  [arXiv:1108.0284 [hep-ph]].


\bibitem{Abbas:2010ns} 
  G.~Abbas, B.~Ananthanarayan, I.~Caprini, and I.~Sentitemsu Imsong,
  ``Improving the phenomenology of $K_{\ell 3}$ form factors with
  analyticity and unitarity,''
  Phys.\ Rev.\ D {\bf 82}, 094018 (2010)
  [arXiv:1008.0925 [hep-ph]].


\bibitem{Abbas:2010jc} 
  G.~Abbas, B.~Ananthanarayan, I.~Caprini, I.~Sentitemsu Imsong, and
  S.~Ramanan,
  ``Theory of unitarity bounds and low energy form factors,''
  Eur.\ Phys.\ J.\ A {\bf 45}, 389 (2010)
  [arXiv:1004.4257 [hep-ph]].


\bibitem{Ananthanarayan:2011xt} 
  B.~Ananthanarayan, I.~Caprini, and I.~Sentitemsu Imsong,
  ``Implications of the recent high statistics determination of the
  pion electromagnetic form factor in the timelike region,''
  Phys.\ Rev.\ D {\bf 83}, 096002 (2011)
  [arXiv:1102.3299 [hep-ph]].


\bibitem{Ananthanarayan:2012tn} 
  B.~Ananthanarayan, I.~Caprini, and I.~Sentitemsu Imsong,
  ``Spacelike pion form factor from analytic continuation and the
  onset of perturbative QCD,''
  Phys.\ Rev.\ D {\bf 85}, 096006 (2012)
  [arXiv:1203.5398 [hep-ph]].


\bibitem{Ananthanarayan:2013zua} 
  B.~Ananthanarayan, I.~Caprini, D.~Das, and I.~Sentitemsu Imsong,
  ``Two-pion low-energy contribution to the muon $g-2$ with improved
  precision from analyticity and unitarity,''
  Phys.\ Rev.\ D {\bf 89}, 036007 (2014)
  [arXiv:1312.5849 [hep-ph]].


\bibitem{Ananthanarayan:2016mns} 
  B.~Ananthanarayan, I.~Caprini, D.~Das, and I.~Sentitemsu Imsong,
  ``Precise determination of the low-energy hadronic contribution to
  the muon $g-2$ from analyticity and unitarity: An improved analysis,''
  Phys.\ Rev.\ D {\bf 93}, 116007 (2016)
  [arXiv:1605.00202 [hep-ph]].


\bibitem{Hill:2010yb} 
  R.J.~Hill and G.~Paz,
  ``Model independent extraction of the proton charge radius from
  electron scattering,''
  Phys.\ Rev.\ D {\bf 82}, 113005 (2010)
  [arXiv:1008.4619 [hep-ph]].


\bibitem{Ananthanarayan:2014pta} 
  B.~Ananthanarayan, I.~Caprini, and B.~Kubis,
  ``Constraints on the $\mathbf {\omega \pi }$ form factor from
  analyticity and unitarity,''
  Eur.\ Phys.\ J.\ C {\bf 74}, 3209 (2014)
  [arXiv:1410.6276 [hep-ph]].


\bibitem{Caprini:2015wja} 
  I.~Caprini,
  ``Testing the consistency of the $\omega\pi$ transition form factor
  with unitarity and analyticity,''
  Phys.\ Rev.\ D {\bf 92}, 014014 (2015)
  [arXiv:1505.05282 [hep-ph]].


\bibitem{Boyd:1995tg} 
  C.G.~Boyd and R.F.~Lebed,
  ``Improved QCD form-factor constraints and $\Lambda_b \to \Lambda_c
  \ell \bar \nu$,''
  Nucl.\ Phys.\ B {\bf 485}, 275 (1997)
  [hep-ph/9512363].


\bibitem{Grinstein:2015wqa} 
  B.~Grinstein and R.F.~Lebed,
  ``Above-Threshold Poles in Model-Independent Form Factor
  Parametrizations,''
  Phys.\ Rev.\ D {\bf 92}, 116001 (2015)
  [arXiv:1509.04847 [hep-ph]].


\bibitem{Peierls}
  R.E.~Peierls,
  ``Interpretation and properties of propagators,'' in
  {\it Proceedings of the 1954 Glasgow Conference on Nuclear and Meson
  Physics}, edited by E.H.~Bellamy and R.G.~Moorhouse (Pergamon Press,
  New York, 1955).

\bibitem{Barton}
  G.~Barton, {\it Introduction to Dispersion Techniques in Field
  Theory}, W.A.~Benjamin, New York, Amsterdam, 1965.

\bibitem{Fermi:2008zz} 
  E.~Fermi,
  ``Lectures on pions and nucleons,''
  Nuovo Cim.\  {\bf 2}, 17 (1955)
  [Riv.\ Nuovo Cim.\  {\bf 31}, 1 (2008)].


\bibitem{Watson:1954uc} 
  K.M.~Watson,
  ``Some general relations between the photoproduction and scattering
  of $\pi$ mesons,''
  Phys.\ Rev.\  {\bf 95}, 228 (1954).

  
\bibitem{Roy:1971tc} 
  S.M.~Roy,
  ``Exact integral equation for pion pion scattering involving only
  physical region partial waves,''
  Phys.\ Lett.\  {\bf 36B}, 353 (1971).

\bibitem{Micu:1973vp} 
  M.~Micu,
  ``Improved optimal bounds using the Watson theorem,''
  Phys.\ Rev.\ D {\bf 7}, 2136 (1973).



\bibitem{Caprini:2005zr} 
  I.~Caprini, G.~Colangelo, and H.~Leutwyler,
  ``Mass and width of the lowest resonance in QCD,''
  Phys.\ Rev.\ Lett.\  {\bf 96}, 132001 (2006)
  [hep-ph/0512364].


\bibitem{Olive:2016xmw} 
  C.~Patrignani {\it et al.} [Particle Data Group Collaboration],
  ``Review of Particle Physics,''
  Chin.\ Phys.\ C {\bf 40}, 100001 (2016).

\end{thebibliography}
\end{document}